\documentclass[pre,twocolumn]{revtex4}

\usepackage{epsfig}

\usepackage{amssymb}
\usepackage{amsmath}
\usepackage[dvips]{color}
\usepackage{lscape}
\usepackage{longtable}
\usepackage{rotating}
\usepackage{float}

\usepackage{bm}

\newcommand{\be}{\begin{equation}}
\newcommand{\ee}{\end{equation}}
\newcommand{\bea}{\begin{eqnarray}}
\newcommand{\eea}{\end{eqnarray}}

\def\pmb#1{\setbox0=\hbox{#1}
        \kern-.025em\copy0\kern-\wd0
        \kern.05em\copy0\kern-\wd0
        \kern-.025em\raise.0433em\box0 } 

\begin{document}

\title{Fractal dimension of a liquid flows predicted coupling  an Eulerian-Lagrangian 
approach with a Level-Set method}
\author{Paolo Oresta$^1$, Arturo De Risi$^1$, Teresa Donateo$^1$,
and Domenico Laforgia$^1$}
\affiliation{
$^1$ Universit\'a del Salento, 
Dipartimento di Ingegneria dell'innovazione, 73100 Lecce, Italy
}
\date{\today}



\date{\today}

\setcounter{page}{1}

\begin{abstract}
The fractal dimension of a liquid column is a crucial parameter in several models
describing the main features of the primary  break-up occurring at the interface
of a liquid phase  surrounded by the gas-flow. In this work, the deformation of
the liquid phase has been numerically studied.
The gas-phase is computed as a continuum in an Eulerian frame while the liquid
phase is discretized in droplets Lagrangian tracked and coupled via the momentum
equation with the surrounding gas flow.
 The interface is transported by the flow field generated because of the particle
 forcing  and it is numerically computed using the Level-Set method.
  Finally, the fractal dimension of the interface is locally
  estimated and used as criterion for the model of the primary
  breakup.
\end{abstract}

\maketitle


\section{Introduction}

The break-up of a liquid flow leads the evaporation process in a wide range of
applications such as in combustion, exhaust gas post-treatment and other devices.
The rate of break-up enhance the liquid evaporation  since the 
 surface/volume ratio increases.
The atomization occurs according with two main mechanisms: the
primary and the secondary break-up \cite{Wang08}.

Firstly, coherent structures detach from the liquid/gas interface up to break in spherical
droplets because of either the Rayleigh-Taylor or Kelvin-Helmholtz instability.
  The instability is locally modulated by waves depending on the velocity and
  the pressure fluctuations induced on the smaller length scales because of the
  turbulence (primary break-up).   

  Thus, each droplet, occurring in the primary breakup, fragments up to
  evaporate because of the shear stress induced by the surrounding gas-flow
  (secondary breakup).

In the case of low pressure injection, the break-up depends mainly on the surface energy
    while  the interface instability plays a negligible role.
        The surface energy could be estimated using a model based on the fractal dimension of
	    the liquid/gas interface \cite{DeR09}.
The connection between the shape of the liquid column and the atomization
process has been pointed out in several experimental studies using the fractal
dimension \cite{Grout07, Moyne08, Weixing00}. 
The topology of the core region is challenging to investigate using experiments 
since the measurements are affected by the resuspended droplets detaching from
the liquid-gas interface.

Numerical simulation appears to be a promising tool
for this purpose.
In the last two decades there has been progress on this topic improving the
numerical methods for the 
the solution of the governing equations of each phase
and for the
topological  treatment of the fluid-gas interface.
Following, the the most remarkable
    techniques used in literature have been depicted.

  The Front Tracking \cite{Front1}, the Volume of Fluid method
    \cite{Vol1} and the Level-Set \cite{Level1} are the main 
    front-capturing methods.

    The Front Tracking method is based on the Lagrangian treatment of the
    interface that is described by ideal particle (markers).
    The markers are connected each other and they are moved like an ideal fluid
    particle. The connectivity between discrete elements
    covering the interface is not straightforward in three spatial dimensions.
    The drawback of this method depends on the collapsing, the stretching and
    the deformation of some discrete elements. These elements should be
    periodically replaced in order to preserve the  curvature and the mass of the fluid
    confined by the interface.
    These difficulties vanish if the interface is described using methods in the
    Eulerian frame as the Volume of Fluid method or the Level-Set method.
    
    In the Volume of Fluid method each phase is represented by the volume
    fraction of the fluid per each computational cell and the interface evolves
    according to the advection equation.
    Despite of the simplicity in describing the connectivity,
     a thickness of the interface is artificially introduced.
     Thus, the curvature and the surface tension are not well estimated. The
     advantage of this method consists in the accurate mass conservation.
     With the Level-Set method the interface is described  by the iso-contour of
     an implicit function $\phi (\textbf{x})=0$, defined in all the points
     $\textbf{x}$ of a fixed computational domain $\Omega$, whose evolution is governed by the
     advection equation. 
     The interface, the outer region and the inner region are defined by
          $\phi(\textbf{x})=0$, $\phi(\textbf{x})>0$, $\phi(\textbf{x})<0$,
	       respectively.
	            The implicit function
     ensure the description of complex gas-fluid interface 
     without the stringent conditions coming from the parametric representation
     of the surface.
     With the Level-Set method the interface is well defined and the surface
     tension is precisely predicted since depends on the gradient of
     the implicit function, only.

Traditionally, the Level-Set involves the numerical solution in the
Eulerian frame for both the
liquid and gas phases. This method fails for high Reynolds number because 
the steep gradients of density  and viscosity across the interface 
 requires a fine
 computational grid
making  tricky the fast simulation of such flows.
Moreover, this procedure is prohibitive in the industrial applications, such as
engine combustion, because the liquid column evolve in a  wide region
far away from the injector. 

The Eulerian-Lagrangian approach is a  strategy to make faster the simulations preserving a reasonable accuracy
for the interface tracking.
The gas-phase is computed as a continuum in an Eulerian frame while the liquid
phase is discretized in droplets Lagrangian tracked and coupled via the momentum
equation with the surrounding gas flow.
The liquid particle has to be smaller enough to feel all the velocity fluctuations 
of the gas-phase in order to preserve as much as possible the 
instability effects induced on the liquid phase.

Once the velocity field has been computed both for the gas and liquid phases the
interface is transported according to the Level-Set method.
The advection velocity  of the interface is equal to the velocity of the liquid particles along
the interface.
Finally, the fractal dimension of the liquid/gas interface is locally
estimated using the box-counting procedure.

\section{Level-Set method}
\label{sec:levelset}

In three space dimension a closed surface separate the whole domain $\Omega$
into the inside domain $\Omega^-$
and the outside domain $\Omega^+$. The border between the inside domain and the
outside domain is called the interface $\partial\Omega$.
An implicit interface representation defines the interface as the iso-contour of
some function $\phi (\textbf{x})$ with $\textbf{x}\in\Omega$.
 The interface, the outer region and the inner region are defined by
 $\phi(\textbf{x})=0$, $\phi(\textbf{x})>0$, $\phi(\textbf{x})<0$, respectively
 \cite{OsSe}.

The gradient of the implicit function $\nabla\phi$ is perpendicular to the
iso-contours of $\phi$.
 Therefore, $\nabla\phi$ evaluated at the interface is a vector that points in
 the same direction as the local unit (outward) normal $\textbf{n}$
  to the interface. Thus, the unit (outward) normal is:

  \begin{equation}\label{eq.norm}
  \textbf{n}=\frac{\nabla\phi}{|\nabla\phi|}
  \end{equation}

  for points on the interface.

  The mean curvature of the interface $k$ is defined as the divergence of the normal
  $k=(\nabla\cdot \textbf{n})/2$ so that $k>0$ for convex regions, $k<0$ for concave regions, and $k=0$ for a
  plane.
  Using the definition of normal vector we obtain:
  \begin{equation}\label{eq.curv}
  k=\frac{1}{2}\nabla\cdot\left(\frac{\nabla\phi}{|\nabla\phi|}\right)
  \end{equation}

Smoothness of the $\phi$ is a desiderable property especially in  using
numerical approximations.
Signed distance functions are a subset of the implicit functions $\phi$ with the
extra condition of $\nabla\phi (\textbf{x})=1$ enforced:

\begin{equation}
\phi(\textbf{x})=min(\textbf{x}-\textbf{x}_i)\mbox{  $\textbf{x}\in\Omega$,
$\textbf{x}_i\in\partial\Omega$}
\end{equation}
 Under these assumption the normal to the interface is:

\begin{equation}      
\textbf{n}=\nabla\phi
\end{equation}
and the mean curvature is: 
\begin{equation}
k=(\nabla^2\phi)/2.
\end{equation}

The signed distance function turns out to be a good choice, since steep and flat
gradients as well as rapidly changing features are avoided as much
as possible.
     The evolution of the interface is governed by the following equation:
              \begin{equation}\label{eq:advection}
	      \phi_t+\textbf{v}\cdot \nabla\phi=0,
	      \end{equation}
	      where $\textbf{v}$ is the velocity of the liquid phase computed at
	      the interface.
	       Once the interface location is known the  Navier-Stokes equation
	       are solved in the gas-phase and the liquid-phase is Lagrangian
	       tracked.

\section{Numerical method for the gas-phase}
\label{NM-gas}

The integral form of the conservative equations
applied to the control volume $V_0$ with a boundary surface $S_0$ and
the unity normal vector pointing outward to the surface $S_0$ defined by
$\textbf{n}$ is:

\begin{equation}
\int_{V_0}\frac{\partial\rho}{\partial
t}dV+\int_{S_0}\rho\textbf{u}\cdot\textbf{n}dS=0
\end{equation}

and the momentum equation is:

\begin{eqnarray}
\int_{V_0}\frac{\partial\rho\textbf{u}}{\partial t}dV
&+&\int_{S_0}\rho\textbf{u}\textbf{u}\cdot\textbf{n}dS
=\nonumber\\
&=&-\int_{S_0}p\textbf{n}dS
+\int_{S_0}\mu\nabla^2\textbf{u}\cdot\textbf{n}dS\nonumber\\
&&+\int_{V_0}\rho\textbf{g}dV+\int_{V_0}\textbf{f}_{liquid}dV
\end{eqnarray}

where $\rho$,  $\mu$ are the density and the dynamic viscosity, respectively. The
flow velocity is $\textbf{u}$, the gravity vector is $\textbf{g}$
, the flow pressure is $p$, the time is $t$ and the feedback forcing induced by
the liquid phase is $\textbf{f}_{liquid}$.

The numerical solver (KIVA) of the gas-phase equations is based on a finite
volume method. 
Spatial difference approximations are
constructed by the control-volume or integral-balance approach, which largely
preserves
the local conservation properties of the differential equations.
In the finite-volume approximations of KIVA, velocities are located at the
vertices and the scalar quantities are located at cell centers.
Surface and volume integrals are approximated using suitable quadrature
formulae. 
Volume integrals of gradient
 or divergence terms are  converted into surface area integrals using the
  divergence
   theorem. The volume integral of a time derivative maybe related to the
    derivative
     of the integral by means of Reynolds’ transport theorem.  Volume and
     surface
      area integrals
       are  performed under the assumption that the integrands are uniform
       within
        cells or on cell faces. Thus area integrals over surfaces of cells
	become sums
	 over cell faces $i$:

	  \begin{equation}
	   \int_\textbf{S}\textbf{F}d\textbf{A}=\sum_i\textbf{F}d\textbf{A}
	    \end{equation}

The effect of turbulence is take into account with a standard version of
$k-\epsilon$ turbulence model modified to include liquid-turbulence interaction.
Evaporating liquid phase is represented by a discrete-particle technique
The momentum exchange is treated by implicit coupling procedures to avoid the
prohibitively small time steps that would otherwise be necessary. 
Turbulence effects on the droplets are accounted.
When the time step is smaller than the droplet turbulence correlation time, a
fluctuating component is added to the local mean gas velocity when calculating each
particle’s momentum exchange with the gas. 
When the time step exceeds the turbulence 
correlation time, turbulent changes in droplet position and velocity are chosen
randomly from analytically derived probability distributions for these changes. 
The temporal differencing is performed with a first-order scheme. 

\section{Numerical method for the liquid-phase}
\label{NM-liquid}

XXX-stochastic method

The liquid phase has been modelled like a cloud of rigid spherical particles with all the forces applied
on the particle centroid. 
The time history of each particle is 
numerically computed (Lagrangian particle tracking) using a model for the
 momentum equation with drag and gravity force, only \cite{Maxey, Camp}:

\begin{equation}\label{eq:partlib}
m_p\frac{dv}{dt}=\frac{1}{2}C_D\frac{\pi
d_p^2}{4}\rho |\textbf{u}-\textbf{v}|(\textbf{u}-\textbf{v})+m_p\textbf{g}
\end{equation}

in which $m_p$ is the mass, $d_p$ is the diameter and $C_D$ is the drag coefficient.

The slight deviation from the
Stokes flow has been taken into account with the drag correction $C_D$
proposed by Shiller and Naumann \cite{Shi33} for particle with Reynolds number $Re_p < 1000$.

\begin{equation}
C_D=\frac{24}{Re_p}\left(1+\frac{1}{6}Re_p^{2/3}\right)
\hspace{0.1cm};\hspace{0.5cm}Re_p=\frac{\rho d_p|\textbf{u}-\textbf{v}|}{\mu}
\end{equation}

We can rearrange the equation
(Eq. \ref{eq:partlib}) as follows:

\begin{equation}\label{eq:partlib2}
\frac{dv}{dt}=\frac{1}{\tau_p}\left(1+\frac{1}{6}Re_p^{2/3}\right)(\textbf{u}-\textbf{v})+\textbf{g}
,\hspace{0.5cm}\tau_p=\frac{\rho_p d_p^2}{18\mu}
\end{equation}

The particle is selectively sensitive to the flow velocity fluctuations induced
by the turbulence. 
This response of a particle to the surrounding gas velocity field 
 is estimated by the Stokes number $St=\tau /\tau_p$, in which
 $\tau$ is the time scale of the eddy interacting with the particle
 characterized by the relaxation time $\tau_p$.
 The time scale $\tau=L/U$ depends on the characteristics length $L$ and velocity
 $U$ of the gas-phase.

\begin{figure}[!h]
\begin{center}
\includegraphics[width=0.5\textwidth]{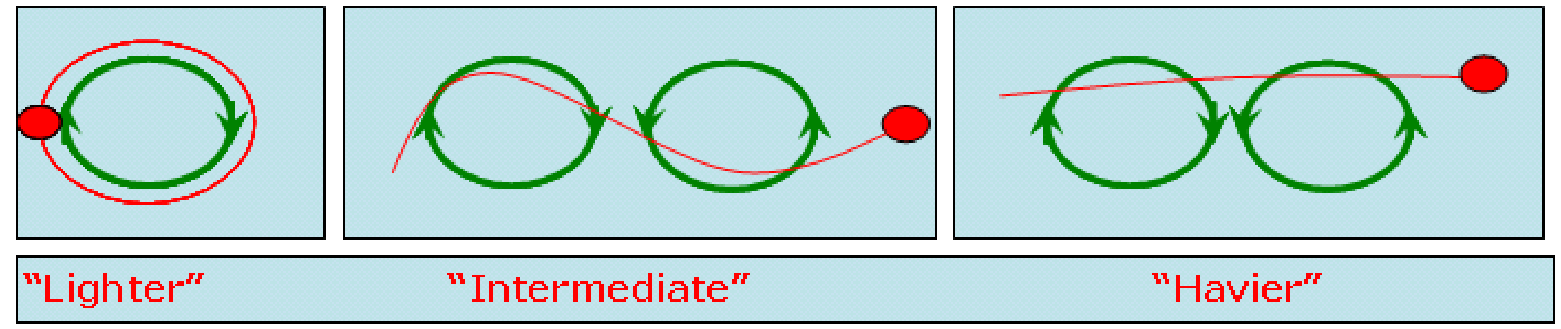}
\caption{\small Qualitatively description of the interaction between
particle and vortices. $St <<1$ - lighter, $St\sim O(1)$ - intermediate, $St>>1$ - heavier.}
\label{fig:interaction}
\end{center}
\end{figure}

Specifically, the droplet with very small
Stokes number will simply be a flow tracers $St<<1$ (lighter particle). For increasing the
Stokes number, the particle trajectory will diverge from the flow stream-line until 
the particle motion will be completely independent from the surrounding flow $St
>> 1$ (heavier particle). If the Stokes number is order one the particle has the
maximum interaction with the velocity fluctuation of the gas-phase \cite{Sq90}.
In Figure \ref{fig:interaction} it is qualitatively depicted the particle-eddy
interaction in which the particle velocity approaches quickly to the fluid
velocity for decreasing the Stokes number. 

\begin{figure}	
  \begin{center}	
   \includegraphics[width=.5\textwidth,angle=0]{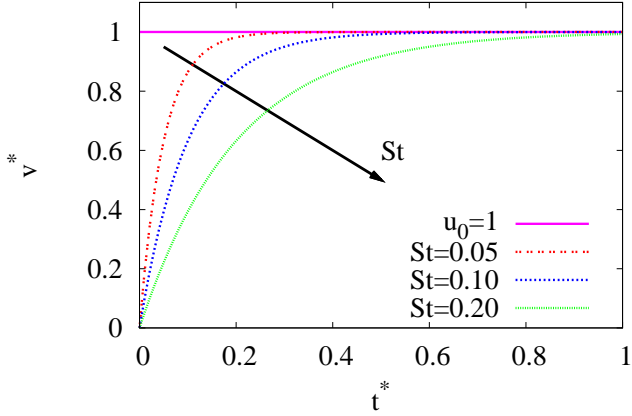}
   \caption{\small Parametric solution of Eq. \ref{eq:sol} for $u^*_0=1$ and three
   Stokes  numbers: $St=0.05$, $St=0.1$, $St=0.2$}
   \label{fig:resp}
   \end{center}
   \end{figure}

Particle behavior depending on the Stokes number can analytically depicted in
the limit of one-dimensional Stokes flow ($Re_p <<1$), neglecting the gravity
for a particle dispersed in a gas-phase with constant velocity $u=u_0$.
Under this hypothesis the equation \ref{eq:partlib2} in dimensionless form read:

   \begin{equation}\label{eq:partlib4}
   \frac{dv^*}{dt^*}=\frac{1}{St}\left(u^*_0-v^*\right)
   \end{equation}

   where $u^*_0=u_0/U$, $v^*=v/U$ and $t^*=tU/L$.
   The solution of the previous equation (Eq. \ref{eq:sol}), in the case of initial particle
   velocity equal to zero, has been plotted in Fig. \ref{fig:resp}.

   \begin{equation}\label{eq:sol}
   v^*=u^*_0\left(1-e^{-t^*/St}\right)
   \end{equation}

In this scenario, the particle diameter, in the Lagrangian model of the liquid
phase, should be small enough in order to make each particle sensitive to the
velocity fluctuation of the smallest vortical scale. 
In the turbulent regime the smallest vortical scale is defined by the
dissipative length scale $\eta$, the velocity $U_k$ and the time $\tau_k$.
Following the Kolmogorov theory the large scale ($U$, $L$, $\tau$) and the
dissipative scale ($U_k$, $L_k$, $\tau_k$) are 
connected according to the following relations:
\begin{equation}\label{viscous0}
 \eta=\left(\frac{\nu^3}{\epsilon}\right)^{1/4},\hspace{0.5cm}U_{k}=(\nu\epsilon)^{1/4}\hspace{0.5cm}\tau_{k}=\left(\frac{\nu}{\epsilon}\right)^{1/2};
   \end{equation}

in which $\epsilon=U^3/L$ assess the energy in the whole domain.
Thus, the equation \ref{viscous0} read:

     \begin{equation}\label{viscous1}
      \frac{\eta}{L}=Re^{-3/4},\hspace{0.5cm}\frac{U_{k}}{U}=Re^{-1/4}\hspace{0.5cm}\frac{\tau_{k}}{\tau}=Re^{-1/2};
       \end{equation}
        where $Re=(UL)/\nu$ is the Reynolds number evaluated at the large scale.
	 
	   The Stokes number  based on the Kolmogorov scales
	     $St_k$ is: 

	         \begin{equation}\label{Stk}
		   St_k=\frac{\tau_p}{\tau_k}=\frac{\tau_p}{\tau}\frac{\tau}{\tau_k}=StRe^{1/2}
		     \end{equation}

Thus, if $St_k<<1$ each particle of the liquid phase takes into account the
instability effects occurring at the liquid/gas interface since the liquid particle is 
sensitive to the velocity fluctuation in the whole range of the turbulent
scales occurring in the computational domain.
For this reason, the particle diameter should be satisfy the equation \ref{Stk2}:

$$
 St_k<<1\Rightarrow StRe^{1/2}<<1\Rightarrow  
 $$

\begin{equation}\label{Stk2}
 \Rightarrow d_p <<3\sqrt{2}\cdot L\cdot\sqrt{\frac{\rho}{\rho_p}}\cdot Re^{-3/4} 
  \end{equation}

\section{Results}\label{sec:results}

The time and the spatial evolution of the gas/liquid interface has been estimated using the Level-Set method. 
According to the Level-Set method, the evolution of the interface is governed by the advection equation \ref{eq:advection}. 
The initial condition of Eq. \ref{eq:advection} is $R_s=d_i/2$, in which $R_s$ is the radius of curvature of the liquid phase
and $d_i/2$ is the radius of the injector. 
The liquid particles are injected into the gas phase within a prescribed maximum angle $\alpha_i$.
In order to enforce both the angle $\alpha_i$ and the injector diameter $d_i$ the particles have been dispersed at 
the following distance from the orifice: 

\begin{equation}
h_i=\frac{d_i}{2}tg\left(\frac{\alpha}{2}\right)
\end{equation}

The center of the sphere, with radius $R_s$, is defined in a Cartesian frame by $(x_s,y_s,z_s)$, where $y_s=y_i$ and
$x_s=x_i$, while $z_s$ depends on $R_s$ and $\alpha_i$ as follows:

\begin{equation}
z_s=z_i-\sqrt{R_s^2-\frac{d_i^2}{4}}+h_i
\end{equation}

\begin{equation}
z_s=z_i-\sqrt{R_s^2-\frac{d_i^2}{4}}+\frac{d_i}{2}tg\left(\frac{\alpha_i}{2}\right)
\end{equation}

\begin{figure}[!h]
\begin{center}
\includegraphics[width=0.5\textwidth,angle=0]{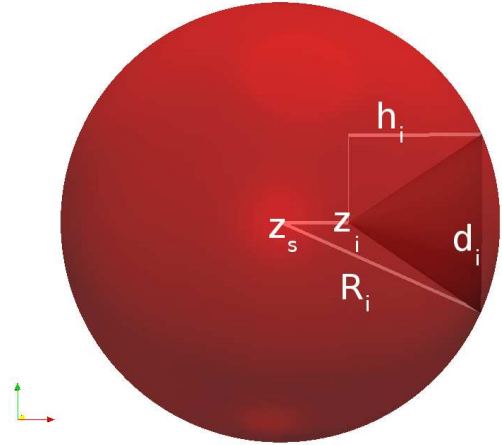}
\caption{\small Initial level-set iso-contour.}
\label{fig:spray.png}
\end{center}
\end{figure}

The numerical simulations have been carried out using an injector with diameter 
$d_i=1.58\cdot10^{-4}$ $[m]$ crossed by a liquid with velocity $U=27$
$[m/s]$ and density $\rho_p=10^3$ $[Kg/m^3]$. 
For such flow the Reynolds number is equal to $Re=U\cdot d_i/\mu=4.3 10^3$. 
According to equation \ref{Stk2} the particle diameter should be  $d_p << 6.9\cdot 10^{-9}$ $[m]$, 
thus the liquid jet has been discretized using droplets with diameter $d_p=10^{-9}$ $[m]$.

\begin{figure}[!h]
\begin{center}
\includegraphics[width=0.2\textwidth,angle=0]{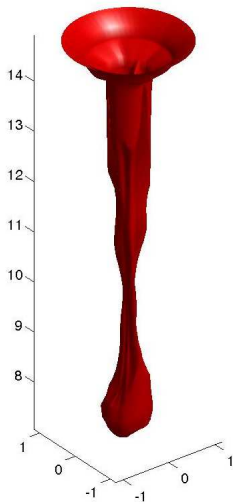}
\caption{\small Interface of the liquid jet.}
\label{fig:fractal.png}
\end{center}
\end{figure}

The liquid jet is organized in an elongated structure 
with inhomogeneous shape modulated by the velocity 
fluctuations leading the instability at the liquid/gas interface. 
The degree of complexity of the interface has been measured using the fractal dimension.
The fractal dimension is the control parameter in several breakup models
since it is the integral measure of the distortion of the liquid jet \cite{DeR09}.
The fractal dimension can be regarded as the general case of the Eulerian dimension.
If we take an object residing in Euclidean dimension $D$ and reduce its linear
size by $1/r$ in each spatial direction, its measure (length, area, or volume)
would increase to $N=r^D$ times the original. This is pictured in the next
figure.

\begin{figure}[!h]
\begin{center}
\includegraphics[width=0.5\textwidth,angle=0]{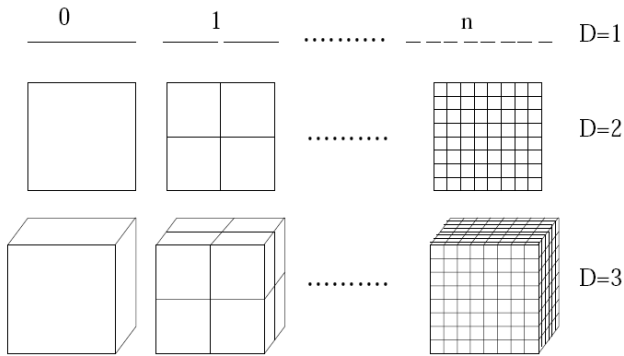}
\caption{\small Examples of figures with integer dimension. Lecture notes of
Fluid Dynamics by R. Verzicco http://ime115.poliba.it/verzicco/}
\label{fig:jet.png}
\end{center}
\end{figure}

We consider $N=r^D$, take the log of both sides, and get
\begin{equation}
log(N) = D log(r)
\end{equation}
If we solve for D.
\begin{equation}
D = log(N)/log(r)
\end{equation}

In the Eulerian frame $D$ is an integer number. Fractals, which are irregular geometric
objects, have a fractional dimension. 
 This theoretical criterion is not straightforward to compute the fractal dimension in numerical computation.
 For this reason, the computation of the fractal dimension of the liquid jet is based on the box-counting procedure.
To calculate the fractal dimension using the box-counting algorithm the liquid/gas interface has been placed on a grid. 
 Then, the number of blocks crossed by the surface has been stored and iteratively computed for resizing grids.
The fractal dimension $D$ is the slop of the best fit in a log-log scale plane of 
the number of boxes $N(s)$ as function of the size $s$ of the resized grids.
The box-counting method is widely
used since it can measure geometry
that are not self-similar.

\begin{figure}[!h]
\begin{center}
\includegraphics[width=.5\textwidth,angle=0]{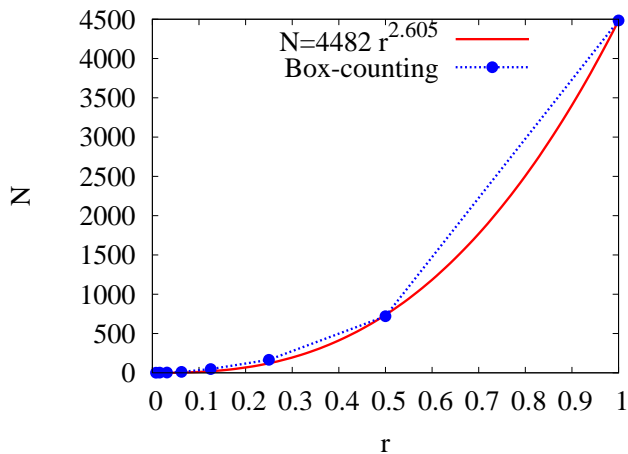}
\caption{\small Best fit of the fractal dimension using the box counting
method.}
\label{fig:box_count_log_2d.jpg}
\end{center}
\end{figure}

The slope of the best fit is equal to $D=2.605$ and it represent the degree of
complexity of the liquid jet surface.

\section{Summary and conclusions}\label{sec:sum}

\end{document}